\documentclass[12pt]{article}
\usepackage{graphicx}
\begin{document}
\title{Now, Factuality and Conditio Humana}
\author{Hartmann R\"omer \\Institute of Physics, University of Freiburg, Germany}
\date{}
\maketitle \pagenumbering{arabic}

\begin{abstract}
The relationship between inner and outer time is discussed. Inner
time is intrinsically future directed and possesses the quality of
a distinguished "now". Both of these qualities get lost in the
operationalized external physical time, which, advancing towards
more fundamental physics, tends to become more similar to space
and even  fade away as a fundamental notion. However, inner time
as a constitutive feature of human existence  holds its place in
the heart of quantum theory and thermodynamics.
\end{abstract}
\section{Introduction}
Time lies next to the hot focus of our mode of existence as
conscious beings and has always been a permanent subject of human
thinking and philosophy. Our world and even our own mind is given
to us in the inexorably temporal form of a movie-like course of
appearances rather than as a simultaneous panoramic picture. All
our reflections about this temporality are caught inside the
genuinely temporal structure of a stream of consciousness. In
fact, trying to escape temporality by an effort of thought is an
extremely delicate task at the verge of paradox and
unthinkability.\\ Over the centuries, much philosophical activity
was devoted to a detailed analysis of the temporal mode of human
existence. Prominent names as Augustine, Kant, Hegel, Bergson and
Heidegger are witnesses of this endeavor. Employing the
terminology of Mc Taggart\cite{McTaggart}, internal existential
time of man is an \textit{A-time}, characterized  by future
directedness and in particular by the existence of the temporal
quality of a distinguished \textit{"now"}. The window of this
"now" moves forward into the future leaving past behind it. The
feeling of this flow is what remains, if all sensory input is
neutralized. According to Heidegger the driving force behind this
motion of the "now" is not a push from the past but rather a pull
from the future originating in the fundamental structural feature
of \textit{"worry"}(German: "Sorge") of human existence.\\ One
particularly important fruit of the persistent philosophical
concern about time is the emergence, clarification and sharpening
of the concept of outer or physical time. Here time has been
tamed, harnessed and subdued to measurability by clocks. External
physical time differs from internal time in several respects.
Again using Mc Taggart's terminology, it can be denoted as a
\textit{B-time}, a scale time representable by the set of points
on a line or a set of real numbers. There is no quality of a
distinguished "now", all time points are equivalent just as points
on the line. Moreover, future directedness is not necessarily
related to physical B-time. The concept of physical time has been
spectacularly successful. Its power manifests itself in the
omnipresence of clocks which hold all of us under their sway.
There is a common tendency to accept physical B-time as the only
exact and in a sense only real notion of time and to reduce
internal A-time to a subordinate or even illusionary status.
Investigating the relationship between internal and physical time
will be the main subject of this note.\\
\section{"Spacialization" and evaporation of physical time}
Along with the advancement towards more fundamental levels of
physics, physical B-time shows a tendency to become more and more
similar to space. In addition, there are strong indications, that,
together with space, it might disappear altogether as a
fundamental notion, when pushed to the extremes of cosmology and
quantum
gravity.\\
From the outset,  the lack of a privileged "now" in B-time
parallels the absence of a privileged "here" in space. In this
respect, B-time is more similar to space than A-time. This opens
up the possibility to represent processes in time in a
diagrammatic atemporal way by introducing a spacial time axis.
\\Let us first look at the motion of a point particle. The presence
of this particle in a point x at (or very close to) time t is an
example of a \textit{point event} with a precise localization in
time and space, which can be described by four coordinates, one
temporal coordinate and three spacial coordinates. Then the motion
of the point particle is completely described by its \textit{world
line}, a one-dimensional set of point events giving the position
of the particle at every time
 (see figure 1).
\begin{figure}
  \includegraphics[width=10cm]{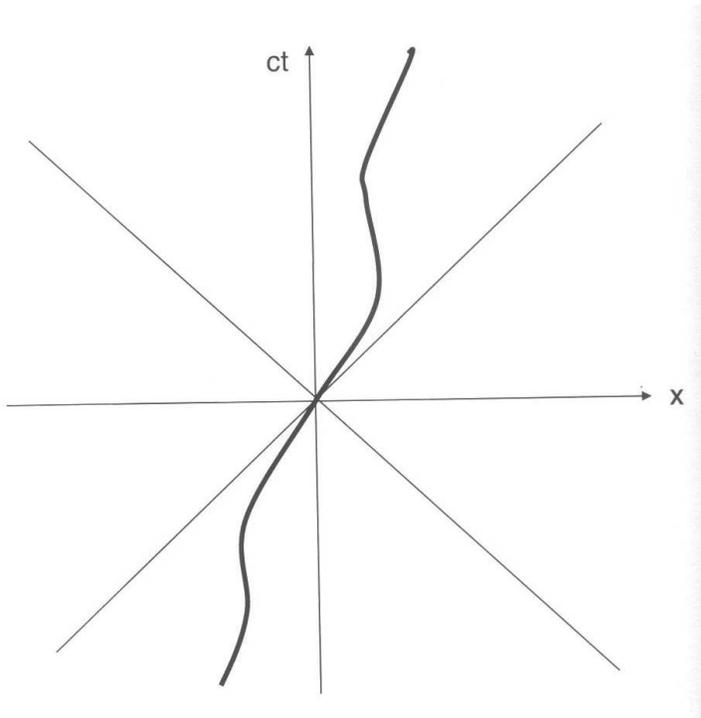}\\
  \caption{Motion of a point particle in the timeless world line representation.
  Two space dimensions are suppressed.}
\end{figure}
 This world line, as it stands, is a line in space, for
instance on a sheet of paper and as such entirely timeless.
Reference to time is only given by its interpretation as a
representation of a motion.
\\This possibility of a timeless representation of processes is
not restricted to physical motions. For instance, the phylogeny of
man can be adequately represented by an totally timeless family
tree (see Figure 2).
\begin{figure}
  \includegraphics[width=13cm]{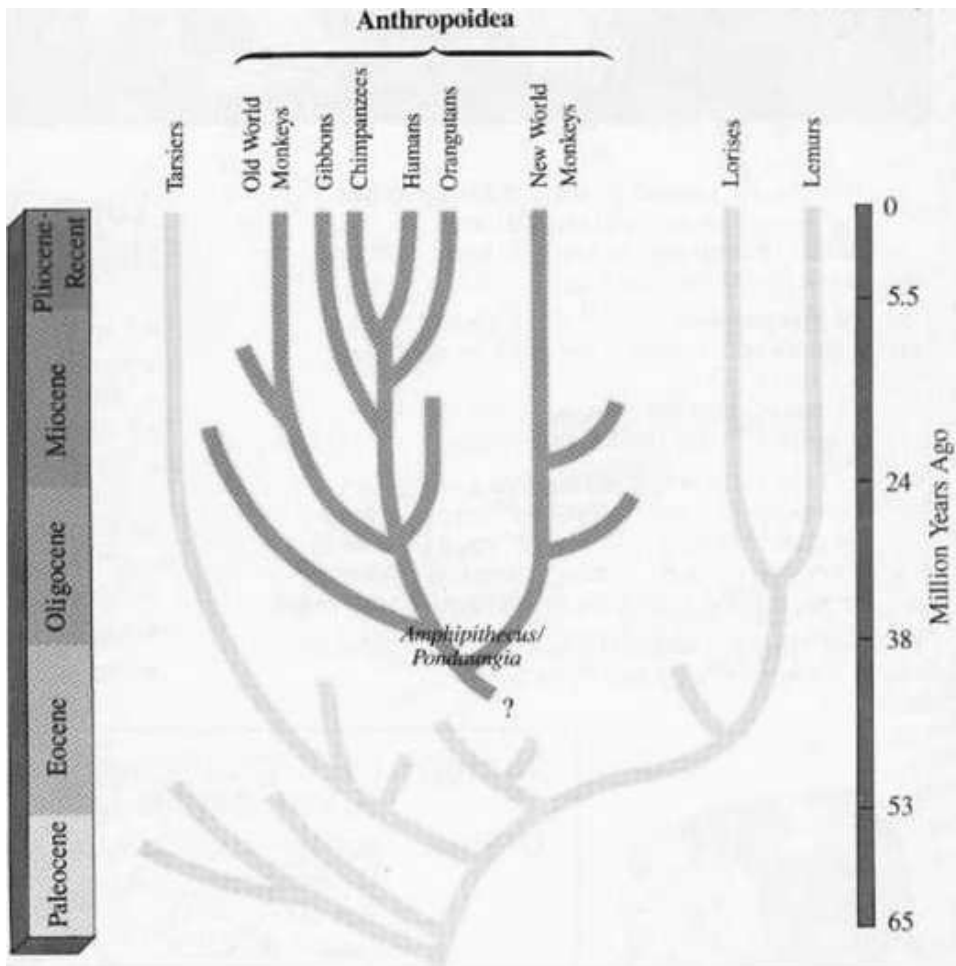}\\
  \caption{Development of hominides as a timeless family tree}
\end{figure}
There is no evident way to decide, which
representation is more correct or "real".\\
Passing from Newtonean space-time to space-time in Special and
General Relativity, one observes that time becomes more and
more similar to space. \\
From the perspective of Kant's philosophy this is not a complete
surprise: According to Kant time and space are forms of intuition
(German: "Anschauungsformen"), through which everything has to
pass, which reaches our mind.  Time is the form of the inner sense
and space the form of the outer sense. (The Kantian distinction
between inner and outer sense is akin to the Cartesian distinction
between res cogitans and res extensa.) As physics is concerned
with the outer world, externalized physical time can be expected
to become more similar to space. \\
Let us now follow the increasing "spacialization" of time from
Newtonean mechanics to General Relativity Theory in more detail.\\
 a) In Newtonean space time the
transition between the coordinates of point events in different
inertial systems is performed by Galilei-transformations. It is
possible to define a global Newtonean world time attributing a
time coordinate to every point event such that time differences
$\tau_{i,j}=t_i -t_j$ between any two point events are independent
of the inertial system. This implies in particular, that the
simultaneity $\tau_{i,j}=0$ of two events is an invariant notion.
Moreover, one immediately obtains
\begin{equation} \label{timedifference}
\tau_{1,2}+\tau_{2,3}=\tau_{1,3}.
\end{equation}
Space behaves differently under Galilei-transformations. Spacial
coincidence is a relative, system dependent notion: Point events
occurring at the same space point in one inertial frame will in
general occur at different space points in another inertial frame.
Only for \textit{simultaneous} events the spacial distance
$r_{i,j}= |\vec{x}_i-\vec{x}_j|$ has an invariant meaning. Instead
of equation (\ref{timedifference}) one has the \textit{triangular
inequality}
\begin{equation} \label{triangular}
r_{1,2}+r_{2,3}\geq r_{1,3}.
\end{equation}\\
b) In Special Relativity Theory the transition between different
inertial systems is performed by proper orthochronous
Poincar\'e-transformations. Time differences are no longer
invariant but depend on the inertial system just like spacial
distances. In particular, simultaneity becomes a relative, system
dependent notion. To be more concrete, the situation in Special
Relativity
Theory is as follows:\\
Light propagates with the same velocity $c$ in all inertial
systems, and this velocity $c$ is the highest possible velocity
for any signal. Two different point events $e_i$ and $e_j$ are
called \textit{relative timelike}, if a subluminar signal can be
exchanged between them, \textit{relative lightlike}, if only a
luminal signal can be exchanged between them and \textit{relative
spacelike}, if they cannot be connected by a signal. It turns out,
that the temporal order of relative timelike or lightlike signals
is independent of the inertial systems. In other words, the sign
(but not magnitude) of the time coordinate difference of such
signals is invariant. This assures that the temporal order of
cause and effect is the same in every inertial system as it
should. On the other hand, the temporal order of relative
spacelike events is system dependent.  It is always possible to
change the sign of time coordinate differences of such events by a
change of the inertial system, and there is always an inertial
system, in which relative spacelike events are simultaneous. ( Of
course simultaneous events are relative spacelike, because
evidently a
signal connecting them would have to have infinite velocity.)\\
For relative timelike events $e_i$ and $e_j$ there is an invariant
measure of their temporal distance: $\tau_{i,j}$ is defined as the
so-called \textit{proper time} measured along a geodesic, i.e.
straight world line connecting $e_i$ and $e_j$. If $e_2$ lies in
the future of $e_1$ and $e_3$ in the future of $e_2$, then rather
than the equality (\ref{timedifference}) an inequality
\begin{equation} \label{timetriangular}
\tau_{1,2}+\tau_{2,3}\leq \tau_{1,3}.
\end{equation}
holds, a triangular inequality similar to the spacial inequality
(\ref{triangular}) but with a $\leq$ instead of a $\geq$ sign.
(This $\leq$ sign is the origin of the famous twin paradox: The
proper time difference for a worldline connecting two relative
timelike events is maximal for a geodesic.) For relative lightlike
events $\tau_{i,j}=0$, and (\ref{timetriangular}) remains valid,
if we allow some of the proper time differences in it to vanish.\\
For relative spacelike events $e_i$ and $e_j$ it is possible to
define an invariant spacial distance $r_{i,j}$, which is simply
the spatial distance in an inertial system in which $e_i$ and
$e_j$ are simultaneous. The spacial triangular inequality
(\ref{triangular}) remains valid as long as $e_1$, $e_2$ and $e_3$
lie in one spacelike plane or, equivalently, if all three events
are simultaneous in an appropriate inertial system. So, we notice
yet another similarity between time and space in Special
Relativity Theory: Both time-
and space- differences fulfill triangular inequalities.\\
c) The unification of space and time goes even further in General
Relativity Theory. Global inertial coordinate systems are no
longer definable, but an observer, freely falling in a
gravitational field, can at least realize a \textit{local inertial
system}, in which the laws of Special Relativity are valid for
small deviations from the origin of the space time coordinates.
This is just a reformulation of Einstein's \textit{equivalence
principle}. The concept of signals and the distinction between
relative timelike, lightlike and spacelike events remains valid.
Space and time are fused in the geometric concept of a
pseudoriemannian space-time manifold $M$ and point events are
points in $M$.  On $M$ a metric tensor $g_{\mu \nu}$ is defined,
which allows to measure the length of world lines in $M$. Also the
concepts of geodesics and geodesic distances remain well defined
and the inequalities (\ref{triangular}) and (\ref{timetriangular})
still hold (at least if $e_1$, $e_2$ and $e_3$ are not too far
apart). Timelike and lightlike geodesics are the worldlines of
massive
particles and photons respectively in a gravitational field.\\
In sharp contrast to Newtonean physics and Special Relativity,
space-time is no longer the external arena on which the play of
physics is performed, but the arena takes part in the play: The
curvature of the space-time metric, which measures its deviation
from the "flat" geometry of Special Relativity depends on the
distribution of energy and momentum of matter in the physical
universe. The view of the universe as suggested by General
Relativity is a timeless geometric structure often referred to as
a \textit{block universe}. The dynamics of particle motions
appears to be frozen in a concept of world lines in the space-time
manifold $M$, which does not suggest any natural definition of a
time coordinate. Also fields like the electromagnetic field or the
gravitational field appear in the timeless geometric mathematical
form of sections in certain
bundles over the space-time manifold $M$.\\
Physical B-time, as opposed to internal A-time is not necessarily
directed. In fact, neither Newtonean time nor time in Special
Relativity nor the concept of pseudoriemannian space-time
manifolds in General Relativity is connected to any notion of a
preferred time direction. Moreover, the fundamental laws of
physics are invariant under time inversion. A tiny violation of
time inversion invariance in weak interactions may look like a
possible exception, but this asymmetry is probably due to
spontaneous symmetry breaking: It results not from an asymmetry of
the fundamental equations of physics but from an asymmetry of
their solution. This also applies to the apparent asymmetry
imposed by the space-time metric of the expanding universe, which
is a time asymmetric solution of the symmetric Einstein equations
of General Relativity. The  directedness of time coming from the
second law of thermodynamics will be discussed later.\\
One of the virtues of General Relativity Theory is that it is able
to predict borders of its validity. The singularity theorems of
Hawking and Penrose prove that certain world lines will inevitably
terminate in space-time singularities, where curvature quantities
become infinite. This, for instance, occurs, when black holes are
formed. Also, all time- or lightlike worldlines of the visual
parts of our universe originate in a gigantic "big bang"
singularity. No time parameter can be extended before this
singularity. Moreover, adding some elements of quantum theory, one
sees that the "spacialization" of time turns into an abdication
and successive loss of both time and space. In fact, time and
space lose their meaning for distances of the order of Planck's
time $t_P=\sqrt{\hbar G/c^5} \approx 5.4\cdot 10^{-44} s $ or
Planck's length $l_P=c t_P =\sqrt{\hbar G/c^3} \approx 1.6\cdot
10^{-35} m$, where $G$ is the gravitational constant and $\hbar$
Planck's quantum of action. This can be seen in the following way:
In order to resolve spacial distances $l$ by a measurement,
according to Heisenberg's uncertainty principle one has to provide
momentum of order $\hbar /l$ in a volume $l^3$. By Einstein's
equations of General Relativity this gives rise to a change of
space-time curvature characterized by a curvature radius $L$ with
$1/L^2=Gp/l^3 c^3=\hbar G/l^4 c^3$. The measurement of $l$
certainly becomes ill-defined if the resulting curvature change
$L$ is of the order of magnitude of $l$. Equating
$l$ and $L$ gives $l=L=l_p=\sqrt{\hbar G/c^3}$.\\
As a consequence, the classical General Relativity Theory becomes
unreliable in a neighborhood of order $l_P$ or $t_P$ of the
singularities predicted by this theory, including the big bang singularity.\\
Although quantum field theory on fixed pseudoriemannian space-time
manifolds is by now well developed, quantum theory of space-time
itself is still in its infancy. The preceding argument shows that
quantum effects become dominant at distances of the order of
magnitude of Planck's length and time. Also the distinction
between timelike, spacelike and lightlike separation becomes
blurred by quantum indeterminacies. The problem of finding a
Quantum Gravity Theory\cite{Rovelli2008}, i.e. a quantum theory of
space-time is unsolved and perhaps the greatest challenge of
fundamental theoretical physics. Describing, comparing and
assessing the various current approaches towards this problem
would at least require a book and be way beyond the scope of this
note. We restrict ourself to some short
remarks about three different approaches.\\
a) \textit{String Theory}\cite{StringTheory2007} starts out from
quantizing the motion of a  one-dimensional string like object
vibrating in a background  space-time manifold $B$. Planck's
length is closely related to the tension of this string. In fact,
the string has to be supersymmetric and carries spin degrees of
freedom. There are good arguments that the quantization of the
string leads to a massless particle of spin 2 to be identified
with the quantum of the gravitational field. (However, the
emergence of curved four-dimensional space-time is not quite
clear.) In addition, the claim is, that  String Theory describes
all particles and interactions of physics. Later on, it turned
out, that for the sake of consistency the quantum theory of the
string should also contain higher dimensional vibrating objects
called \textit{branes}. The technical and, more seriously,
conceptual difficulties of String Theory are formidable. So far,
no precise formulation of its theoretical framework is available.
It is not yet clear how to get rid of the undesirable background
dependence of string theory. For consistency, the background
manifold $B$ must be ten-dimensional, and six space dimensions
must be "compactified" i. e. curled up at a small scale in order
not to be directly visible. The precise form of this
compactification is undetermined and seems to be highly arbitrary.
(According to rough estimates there are more than $10^{500}$
different compactifications.) The particle content and the
interactions of the resulting theory radically depend on the
compactification, leading to a loss of predictive power as long as
no principle for identifying preferred compactifications can be
given. To make things worse, no known compactification reproduces
the well established standard model of
elementary particle theory.\\
b) \textit{Loop Quantum Gravity}\cite{Thiemann2007, NPZ} belongs
to a class of theories which assume, that for very small scales of
the order of magnitude of $l_P$ space-time or space loses its
continuous manifold character and becomes discontinuous and
discrete like a lattice. On larger scales the number of lattice
points should become so large that in the average a
\textit{continuum limit} is a good approximation, such that a
smooth manifold structure is regained. So, the smooth space time
of Newtonean theory and Special and General Relativity is
conceived as a derived, approximate notion. The virtue of such
theories is that singularities at small scale are excluded from
the outset. However, it is still not clear, how and under what
conditions the smooth continuum limit is
obtained.\\
c) \textit{Canonical Quantum Gravity} tries to apply the procedure
of canonical quantization to Einstein's theory of gravitation. In
its most popular form, it leads to the \textit{Wheeler- de Witt
equation}\cite{Kiefer2000}, an equation of the structure
\begin{equation} \label{wheeler}
H(g^{(3)},\delta/\delta g^{(3)})\Psi(g^{(3)})=0 .
\end{equation}
Here, the "Schr\"odinger wave function" $\Psi(g^{(3)})$ is a
functional depending on metric $g^{(3)}$ on a three dimensional
manifold, and the Hamiltonean $H(g^{(3)},\delta/\delta g^{(3)})$
depends on $g^{(3)}$ and the functional derivatives with respect
to $g^{(3)}$. Unlike a normal Schr\"odinger equation in quantum
mechanics, the Wheeler- de Witt equation contains no time
derivative of the wave function and is, in this sense, completely
timeless. This is related to the absence of any natural time
coordinate in General Relativity. Something like time may be
recuperated in a quasiclassical approximation (After all, the
space-time we are living in is classical) of the type
\begin{equation}
\Psi\sim \exp(iS/ \hbar) .
\end{equation}
In this case, the support of the action functional $S$ is
decomposed into one-dimensional families of 3-metrics, which can
be interpreted as metrics on 4-dimensional manifolds. Depending on
the solution $\Psi$, the signature of the 4-metric may be
pseudoeuclidean, and the parameter along such a one-dimensional
family can be interpreted as a time coordinate. The Big Bang
singularity may appear in the form of a boundary condition on
$\Psi $. We see, that also in this approach, time only appears as
a secondary, derived and approximate notion. In addition, there
are also solutions of the Wheeler- de Witt equation which do not
allow for the introduction of a time coordinate. It must also be
admitted that the precise  mathematical definition of the
Hamiltonean and the wave functional in the
Wheeler- de Witt equation is not yet clear.\\
These examples for Quantum Gravity theories clearly show, that the
tendency of a successive abdication of physical time is continued,
if quantum effects on space-time are taken into account.
\section{The revival of A-time in quantum theory}
The progressive spacialization of physical time becomes also
evident in a comparison of nonrelativistic quantum mechanics,
quantum field theory on Minkowskian space and quantum field theory
on curved background space-time manifolds.\\
Nonrelativistic quantum mechanics is the quantum version of
Newtonean mechanics. Space and time play quite different roles:
Whereas space is represented by position operators acting on the
Hilbertspace of quantum states, time is present as a classical
parameter $t$ of B-type. In the Schr\"odinger picture of quantum
mechanics, the state vector $\psi(t)$ depends on this parameter.
Alternatively, in the Heisenberg picture all observables depend on
$t$.\\
In quantum field theory the Heisenberg picture is conceptually
favored and both space and time are reduced to the same status of
classical parameters such that the field operators
$\Phi(t,\vec{x})$ are functions of $t$ and $\vec{x}$. It is still
formally  possible to construct spacial and temporal localization
operators from these field operators\cite{BF2002b, BFR2010} but
with a lot of arbitrariness. The similarity of the time and space
parameters is further enhanced in special relativistic quantum
field theory and even more so in quantum field theory on curved
background manifolds. The parametric space time dependence is of
paramount importance for the axiomatic formulation of local
quantum field theory\cite{Haag}. The concept of locality has to be
revised, if also the geometry of time and space is subject to
quantization and loses its fundamental meaning as we saw in the
preceding paragraph.\\
In view of the spacial character of physical (B-)time and its
eventual disappearance, time is sometimes assumed to be unreal and
illusionary.  Attributing such an ontologically subordinate status
to time signals a strong physicalistic and  reductive attitude
connected to the claim that, at least in principle, everything
should be describable in terms of physics. We shall argue that
this creed is of low plausibility. Right here we should notice
that physics is a ´highly developed, sophisticated and
spectacularly successful method to build up a mathematical model
of the world. But by definition a model concentrates on features
amenable to its framework, and it is certainly a methodological
mistake to identify a model with what it modelizes. \\
Classical physics silently supposes an external observer merely
registering  phenomena of the observed system without influencing
it in an essential and non-negligible way. This supposition is no
longer tenable in quantum theory, where measurement will in
general change even pure states. The role of the observer becomes
an active rather than a merely registrating one, and time in the
form of the A-time of the observer has its place right in the
heart of
quantum theory:\\
A measurement result is factual and appears via the "now" of the
observer. The directedness of the observer's A-time is encoded in
the non commutative structure of quantum theory. The order of
successive measurements is vital, and in general only the result
of the last and latest measurement can claim factual status,
whereas the factuality of previous measurement results is
destroyed by subsequent measurements. Notice, that the composition
$P_1P_2$ of two proposition observables $P_1$ and $P_2$ means that
$P_1$ is applied and measured \textit{after}$P_2$.\\
A complete physical description of a measurement process in terms
of quantum dynamics is not available. In quantum dynamics, time
development is given by unitary transformations of states or
observables. A measurement results in a non unitary reduction of
the quantum state. Of course, a measurement process is accompanied
by a quantum physical process, but this physical description does
not seem to be exhaustive. In fact, no clear purely physical
criterium lends itself to qualify a physical process as a
measurement process. In addition to being a physical event, a
measurement is also an act of cognition. In the subtle role of the
measurement process, quantum physics has an open door to cognition
and  epistemology, and such philosophical considerations cannot be
ignored as easily as in classical physics. Unitary time
development of quantum dynamics is completely deterministic, and
nondeterministic randomness enters only through measurement. Also
the causal closure of pure quantum dynamics is broken by the
measurement process and the freedom of a choice of the observable
to be measured. (This also applies to classical theory but becomes
more conspicuous in quantum theory.)\\
The thermodynamic time arrow pointing towards an increase of
entropy is another case, where an inroad of directed A-time into
physics can be observed. The notion of entropy rests on an
incomplete description of a system in terms of \textit{macroscopic
states}, where a complete description in terms of
\textit{microscopic states} is unfeasible. The transition from a
microscopic to a macroscopic description is performed by the
application of a \textit{coarse graining} procedure, which
associates a certain well-defined mixed statistical product state
to the system and thereby a definite value to its entropy. Normal
microscopic physical dynamics (unitary in quantum theory,
symplectic in classical theory) does not change the entropy of
this state. An increase of entropy only results when coarse
graining is again applied to the time developed state. Thereby
inaccessible correlation information between the different
subsystems corresponding to the coarse graining is discarded and
entropy is increased. So far, the reasoning was still time
symmetric, because the microscopic physical time development could
have been taken in both time directions. The directedness of
A-time enters, because an observer can only \textit{first}
register a macroscopic state and \textit{then}
discard correlation information and not vice versa.\\
For an investigation of the relationship between inner A-time and
outer physical B-time we need a comprehensive framework including
both mental and material systems. \textit{Generalized Quantum
Theory}(GQT) \cite{ARW, AFR2006, FR2010} provides a general system
theory, which arose from physical quantum theory in its algebraic
form leaving out those features which only pertain to physical
systems\footnote{Originally it was  called  "Weak Quantum Theory",
but this led to misunderstanding by non mathematicians.
Admittedly, the term "Generalized Quantum Theory" is  somewhat
unspecific and equivocal and has been used in different senses by
other authors.}. The resulting formalism is still rich enough to
allow a controlled and formally well defined application of
quantum theoretical notions like complementarity and entanglement
far beyond the realm of physics. Mental systems, in particular the
human mind as seen from an inner first person perspective are well
inside the scope of GQT and are even of paradigmatic importance in
this framework.\\

\section{Generalized Quantum Theory}
In this section, we give a brief account of the vital structural
features of GQT in order to ease the understanding of our
argumentation in the subsequent sections. For a full account of
GQT we refer to the original publications\cite{ARW, AFR2006,
FR2010}. References for numerous applications of GQT, which have
been worked out in more or less detail, can be found in
\cite{FR2010} and \cite{QuantumConsciousness}. In \cite{AR2012} a
new application of GQT to order effects in questionnaires is
described. GQT takes over  from quantum physics the following
four fundamental notions:\\
\begin{enumerate}
\item \emph{System}: A system is anything which can be (imagined
to be) isolated from the rest of the world and be subject to an
investigation. In the sequel we shall consider systems containing
also conscious individuals. In contradistinction to, e.g.,
classical mechanics the identification of a system is not always a
trivial procedure but sometimes a creative act. In many cases it
is possible to define \emph{subsystems} inside a system.
 \item \emph{State}: A system must have the capacity to reside in
different states without losing its identity as a system. One may
differentiate between \emph{pure states}, which correspond to
maximal possible knowledge of the system and \emph{mixed states}
corresponding to incomplete knowledge. In the most general form of
GQT, the set of states has no underlying Hilbert state structure.
For some applications (see, e.g., \cite{AFR2004, ABFKR, AFR2008,
AR2012}) one may want to enrich the  minimal scheme of GQT, for
instance by adding this  additional structure.
 \item
 \emph{Observable}: An
observable corresponds to a feature of a system, which can be
investigated in a more or less meaningful way. \emph{Global
observables} pertain to the system as a whole, \emph{local
observables} pertain to subsystems.
\item
 \emph{Measurement}: Doing a
measurement of an observable $A$ means performing the
investigation which belongs to the observable $A$ and arriving at
a result $a$, which can claim factual validity. What factual
validity means, depends on the system: Validity of a measurement
result for a system of physics, internal conviction for self
observation, consensus for groups of human beings. The result of
the measurement of $A$ will in general depend on the state $z$ of
the system before the measurement but will not be completely
determined by it. In GQT as well as in physical quantum theory,
the notion of measurement contains an element of idealization,
because measurement is not described as a temporal process, rather
the focus of attention lies on the factual result obtained
eventually.
\end{enumerate}
In addition to these definitions the following structural features
of GQT are of particular importance, generalizing essential
properties of physical quantum theory. To every observable $A$ we
associate its \emph{spectrum}, a set Spec\,$A$, which is just the
set of all possible measurement results of $A$.
Immediately after a measurement of an observable $A$ with result $a$ in Spec\,$%
A$, the system will be in an \emph{eigenstate} $z_a$ of the
observable $A$ with \emph{eigenvalue} $a$. The eigenstate $z_a$ is
a state, for which an immediate repetition of the measurement of
the same observable $A$ will again yield the same result $a$ with
certainty, and after this repeated measurement the system will
still be in the same state $z_a$. This property, which is also
crucial in quantum physics justifies the terminology ``eigenstate
of an observable $A$'' for $z_a$ and ``eigenvalue'' for the result
$a$. We repeat that this is an idealized description of a
measurement process abstracting from its detailed temporal structure.\\
Two observables $A$ and $B$ are called \emph{complementary}, if
the corresponding measurements are not interchangeable. This means
that the state of the system depends on the order in which the
measurement results, say $a$ and $b$, were obtained. If the last
measurement was a measurement of $A$, the system will end up in an
eigenstate $z_a$ of $A$, and if the last measurement was a
measurement of $B$, an eigenstate $z_b$ will result eventually.
For complementary observables $A$ and $B$ there will be at least
some eigenvalue, say $a$, of one of the observables for which no
common eigenstate $z_{ab}$ of both observables exists. This means
that it is not generally possible to ascribe sharp values to the
complementary observables $A$ and $B$, although both of them may
be equally important for the description of the system. This is
the essence of quantum theoretical complementarity which is well
defined also for GQT. \\
Non complementary observables, for which the order of measurement
does not matter, are called \emph{compatible}. After the
measurement of compatible observables $A$ and $B$ with results $a$
and $b$, the system will be in the same common eigenstate $z_{ab}$
of $A$ and $B$ irrespective of the order in which the measurements
were performed.\\
In quantum physics, \emph{entanglement} is normally explained by
the existence of non separable Hilbert space states, which are
linear superpositions of separable tensor product states.  But
entanglement can also be defined in the framework of GQT, which
contains no reference to a Hilbert space of states \cite{ARW,
AFR2006, FR2010, Roemer2011}. It may and will show up under the
following
conditions:\\
\begin{enumerate}
\item Subsystems can be identified within the system such that
local observables $A_i$ pertaining to different subsystems are
compatible.

\item There is a global observable $A$ of the total system, which
is complementary to local observables $A_i$ of the different
subsystems.

\item The system is in an \emph{entangled state}, for instance in
an eigenstate of the above mentioned global observable $A$, which
is not an eigenstate of the local observables $A_i$.
\end{enumerate}

Given these conditions, the measured values of the local
observables will be uncertain because of the complementarity of
the global and the local observables. However, so-called
\emph{entanglement correlations} will be observed between the
measured values of the local observables $A_i$ pertaining to
different subsystems. These correlations are non local and instantaneous.\\
In physical quantum theory, the singlett state of a two-spin
system is a standard example of entanglement. In this case, the
total spin
\begin{equation}
S^2=  (s^{(1)}_1 +s^{(2)}_1)^2 + (s^{(1)}_2 +s^{(2)}_2)^2 +
(s^{(1)}_3 +s^{(2)}_3)^2
\end{equation}
is the global observable, the individual spin observables
$s^{(1)}_3$ and $s^{(2)}_3$ take over the role of the local
observables complementary to the global observable, and the
entangled singulett state is the eigenstate of $S^2$ with
eigenvalue zero. We see how our generalized definition captures
the essentials of
quantum theoretical entanglement.\\
In physical quantum theory it is not difficult to show that
entanglement correlations cannot be used for signal transmission
or controlled causal influences. In order to avoid intervention
paradoxes this must be postulated for GQT \cite{Roemer2004,
LRW2007, Roemer2011}.\\In view of the possibility of entanglement
correlations in GQT the problematic and anything less than trivial
character of the act of identification of a system becomes even
more acute.

\section{Observables, partitions and epistemic cut}
In classical or quantum mechanics, observables like positions and
velocities seem to be given in a very direct and unproblematic way
by the system itself. The example of entropy\footnote{In physical
quantum theory, as oppposed to GQT, entropy is not an observable
in the technical sense.} shows that even for physical systems the
identification of observables need not be so trivial but sometimes
means a major discovery. For very general systems like those
considered in GQT, observables are not so directly given by the
system and read off from it like location and velocity in a
mechanical system. On the contrary, as already suggested by the
name of an ``observable'', the identification of an observable may
be a highly creative act of the observer, which will be
essentially determined by his horizon of questions and
expectations. This marks a decidedly epistemic trait of the notion
of observables in GQT even more so than in quantum physics.
Moreover, the horizon of the observer will change, not the least
as a result of his previous observations adding to the open and
dynamical character of the set of observables. In humanities, this
iterative and constitutive process is sometimes called the
hermeneutic circle.\\
What has just been said about observables also applies to
\emph{partitioning} a system into subsystems. G.
Mahler\cite{Mahler2004} vigorously pointed out that the
identification of subsystems in a complex system may be a highly
creative act. In general, subsystems do not preexist in a
na\"\i{}ve way but are in a sense created in the constitutive act
of their identification. Partitioning may be considered as a
special case of constituting observables, because partitioning is
achieved by means of \emph{partition observables} whose different
values differentiate between the subsystems. Partition observables
and, hence, the associated partitions may be complementary,
resulting in an incompatibility of different partitions. Two such
incompatible partitions cannot be overlaid in order to arrive at a
common refinement of both of them. A simple physical example of
such a situation are partitions according to position or momentum
of a quantum multiparticle system. In physical systems, the
position observable $Q$ is a privileged partition observable,
which differentiates between subsystems by their different
locations. The paramount importance of the position observable $Q$
is reflected in the fact, that the realm of physics is often
identified with the range of applicability of the position
observable. This is reminiscent of Descartes' denotion of the
material world as "res extensae". Indeed, position and locality
play a vital role in physics down to small distances of the order
of the Planck length $l_P$.\\
The first partition, prior to and prerequisite for every act of
measurement or cognition is the split between observer and
observed system. In quantum theory this is referred to as the
\textit{Heisenberg cut}. In the wider framework of GQT we call it
the \textit{epistemic split.} Just as the Heisenberg cut, the
epistemic split is movable but not removable, because every
cognition accessible to us is the cognition of someone about
something. The epistemic cut may be far outside the observer, if a
remote quasar is observed, or run right through a person's mind in
the case of self observation, but it is never absent. It is
conceivable that there are partitions and observables incompatible
with a given or even any epistemic split. Quantum theoretical
uncertainties are closely related to the epistemic split. As
already mentioned above, in the quantum theory of physical
measurement the time development in the large system containing
both observer and observed is completely deterministic.
Stochasticity enters into quantum theory as a result of the
epistemic split into the subsystems "measured system" and
"observer" or "measurement apparatus" and by the subsequent
projection onto the latter system and the interpretation of the
result obtained there as a statement about the measured system. In
quantum physics, there is a symmetry between observer and
observed. Projection onto the measured system yields the same
probability distributions as the above-mentioned projection onto
the measurement device. One may wonder\cite{Roemer2011}, to what
extent this symmetry has a generalization to GQT in the form of a
certain "inside-outside" symmetry.\\
Coming back to observables in quantum physics and GQT, it is
important to stress, that they already presuppose the epistemic
split. Observables neither exclusively pertain to the observed
system nor to the observer, rather they are located astride on the
epistemic split. Related to this, the observer plays an active
role in the constitution of observables, in the choice of the
observables to be measured and in the factual establishment of the
measurement results.\\
One might be tempted to associate observables with properties of
systems, which, like position or velocity, correspond to
substantial entities best expressed by nouns. In \cite{Roemer2006}
we argue, that this would point in the direction of a one-sided
preference for an ontology of substances. Process observables
associated to transitions and changes of a system and typically
better expressible by verbs are of equal importance, and a balance
between substance and process ontology\cite{Whitehead1978,
Rescher2000} seems to be desirable. We shall have to say more
about this in the following section.
\\For what follows we should also keep in mind, that in GQT inner
observables occurring in introspective self observation are fully
legitimate objects of consideration.

\section{The emergence of physical time}
We already mentioned several times, that, as opposed to internal
A-time, external physical B-time lacks an intrinsic directedness
as well as the quality of a distinguished "now". The physics of
time direction\cite{Zeh} endeavors to establish a physical basis
for the evident directedness of time encountered  everywhere in
the world we live in. In other words, a physical mechanism is
sought, which endows B-time with directedness. (The equally
fundamental problem of the missing "now" is normally not dealt
with.) The investigations related to this task proceed in the
following way: First various "arrows of time" are described, in
which time shows directedness, such as the thermodynamic time
arrow for the increase of entropy, the cosmological time arrow for
the expansion of the universe, the retardation time arrow for the
emission of radiation from a spacially localized source or the
evolutionary time arrow for the formation of structures in our
world. Then arguments are given, that the various time arrows are
aligned, i.e. point into the same direction. Finally, one of the
time arrows is identified as primary and a physical mechanism is
proposed to account for it. Usually, either the cosmological or
the thermodynamical time arrow are placed in the pivotal position.
Here, the cosmological time arrow appears in the role of a
boundary condition for cosmological evolution, whereas a complete
formal deduction of the second law of thermodynamics from the laws
of physics is still missing.\\
In these physical approaches to the directedness of time, the
evolutionary time arrow is interpreted as a reflection of the
thermodynamic time arrow.  This is also offered as the explanation
for the directedness of internal A-time, referred to as the
"psychological time arrow" in this context.\\
In this note, we propose to place internal A-time in the primary
position as a categorial constitutive feature of the human mode of
existence, prior to any physical modelling of the world. Physical
B-time would then result secondarily together with a loss of the
qualities of directedness and "now". One evident advantage of this
view is that it is certainly easier to understand a loss of
features than the emergence of completely new features of time.
Anyhow, as explained in Section 2, A-time is present in the heart
of quantum theory and thermodynamics. \\
Our scenario for the constitution of physical B-time has been
described in detail in \cite{Roemer2004}.  GQT offers itself as a
suitable formal framework, because it allows to treat material and
mental systems on an equal footing. (For a related but in many
respects rather different quantum approach to temporal matter-mind
systems see \cite{Primas2003}.) Here we list the most important
steps on this way, referring to \cite{Roemer2004} for details:\\
\begin{enumerate}
\item After an epistemic split, internal time observables $T_i$
arise in subsystems corresponding to individuals and the time
bounded mode of their conscious personal existence. \item The
partial synchronization between different internal time
observables $T_i$ as well as between many "clock" observables
$T_I$ identified in the outer physical world is certainly not an
effect of causal interactions, but an acausal parallelism which
can be described by entanglement correlations in the sense of GQT
between many subsystems. \item External time is successively
sharpened and operationalized in a long and sophisticated process
of optimizing physical clock observables such that the
afore-mentioned entanglement correlations  become as strict as
possible. (For a nice account see \cite{FG2004}.) At present, the
best clocks are given by atomic systems with oscillations between
sharp energy levels. \item In view of what was said in Section 2,
time becomes more and more similar to space in this process of
externalization. After all, space is the form of the exterior
sense. In the course of "spacialization" of time both the notion
of "now" and the future directedness of time get lost, because
space neither has a privileged "here" nor a natural directedness.
\item We saw that, pushed to the extremes, sharpening of the
concept of physical time eventually will lead to its
deconstruction with a loss of its fundamental meaning. So, the
course of events can be summarized in the following way:
establishment of internal time, establishment of physical time
along with a fading of characteristic features of inner time,
deconstruction of physical B-time.
\end{enumerate}
The distinction between substance and process observables
mentioned at the end of Section 5 can be formalized using the time
observables $T_i$ \cite{Roemer2006a, Roemer2006}. A substance
observable $C$ is an observable which commutes with $T_i$: $T_iC=
CT_i$. This means that either $C$ is compatible with a sharp
localization in time or bears no reference to time whatsoever. The
position observable $Q$ or the angles of an triangle are  simple
examples of substance observables. Process observables can be
defined as complementary to $T_i$: $T_iD \neq DT_i$. By
definition, a temporal process has no precise temporal
localization. The complementarity of substance and process
observables was proposed in\cite{Roemer2006a, Roemer2006} as a
resolution of Zeno's paradox: A flying arrow seems to freeze in
its motion, when attention is focused on its momentary position at
any given instant of time: The arrow never occupies more space
than given by its length. The position of the arrow at a given
time is a substance observable, whereas its total flight is
described by a process observable. The incompatibility of process
and substance in this case is similar to the vanishing of the
concept of the orbit in the description of motion in quantum
mechanics.

\section{Conditio humana}
Evidently, the world is never given to us directly but only as it
appears on  our inner screen. This trivial fact, which in
philosophical terminology is just the phenomenal character of the
world, when taken seriously, has far reaching consequences.
Everything we sensually or intellectually conceive of our world is
shaped and conditioned in a categorial way by the mode of our
existence as conscious individuals. Naive realism asserts that the
world appears to us more or less "like it really is". Sometimes
our categorial cognitive structure is compared to a pair of
colored sunglasses, which can be taken off to allow a look at the
real world. But also this optimistic belief underestimates the
inexorable phenomenality of our existence, which must be the
starting point of every reflection about the way we orient
ourselves in our world. In particular, physics cannot lay its own
foundations but has to be aware of the categorial prerequisites
imposed by our cognitional system and our mode of existence. In
this spirit we already mentioned in Section 3 that a measurement
should not entirely be conceived as a physical process but also as
an act of cognition. This also prevents a complete causal closure
of physics. Of course, the physical process accompanying
measurement has to be investigated and consistency with the
possibility of cognition must be guaranteed. A strict physical
reductionism, trying to reduce "everything" to physics, is unaware
of the phenomenal character of the world and, hence,  of its own
foundations. Moreover, as already mentioned, it runs into the
naive methodological mistake to identify the model with what is
modelled. Everetts's many-world interpretation of  quantum theory
illustrates the bizarre consequences of an extreme physicalism in
the interpretation of measurement: The whole universe, conceived
as a purely physical system, is assumed to split into
several branches as the result of each measurement. \\
The main structural features of the phenomenal mode of human
existence have already been mentioned in passing. We briefly
collect them here, a more detailed analysis can be found in
\cite{Roemer2011b}. (Ref. \cite{Prauss2006} contains a
comprehensive and deep discussion.)
\begin{itemize}
\item \textit{The figure of oppositeness}. In every act of
cognition we experience ourself as an observer, different and set
apart from  what we observe. This is sometimes referred to as the
\textit{egocentricity}\cite{TugendhatDeutsch,
TugendhatItalienisch, TugendhatSpanisch} of human existence. The
epistemic cut between observer and observed is never absent.
 \item \textit{Temporality}. Human existence is inescapably temporal in
 the sense of a future-directed A-time with a privileged "now".
 \item \textit{Factuality}. We live in a world of facts rather than
 a world of potentialities. Everything which appears to us, primarily
 touches us in the form of a fact. In particular, the "now"
 carries the imprint of prototypic factuality.
\end{itemize}

These basic existential features are deeply encoded in the
structure of quantum theory and GQT\cite{Roemer2011b}. The
naturalness and, in a way, a priori structure of quantum theory
has been observed by many authors and has, for instance, been
expressed in full clarity by M. Bitbol\cite{Bitbol}.
\begin{itemize}
\item The epistemic cut is present in the very special and
fundamental role attributed to measurement in quantum theory and
GQT. We saw that observables are located right on the epistemic
cut. Standard reductive physicalism ignores the importance of the
observer and the epistemic cut in favor of the outside world. In
this sense, it is as one-sided and implausible as a solipsistic
world view, which ignores the outside in favor of the inside
world. \item In section 3, we saw, how deeply A-time is encoded
into quantum theory, GQT (and thermodynamics). \item Factuality is
intimately related to quantum theoretical measurement, which
basically amounts to a  transition from potentiality to a
measurement result of factual validity.
\end{itemize}
The categorial scheme of human existence is, of course, the
product of a long development. The temporality of primitive
animals is a total subjection to the undivided factuality of a
simple "now". Memory and the possibility of preparing actions open
up the horizon of temporality eventually resulting in a
differentiation between past, present and future. Causality and
personal freedom, which are often considered to be in
contradictory relationship, actually rely on  one another and  are
in fact offshoots of the same root of such a developed and
differentiated temporality. This phylogenetic process is repeated
in quick motion in the ontogenesis of every human individuum.
Related to the unfolding of temporality is an emancipation from
the tight binding to primitive factuality. Free exploration of the
space of possibilities comes into sight with the capacity for
hypothetical and contrafactual thinking. Along with this
emancipation goes a deepening of the epistemic cut. The precise
form of human existence undergoes a process of varied cultural
evolution and also shows large individual differences. Development
goes on: Man is always rebellious against his categorical
limitations. Philosophy, science and arts grant  visions on
timeless structures. Utopianism challenges factuality and
integrative world views embedding man into an comprehensive
universe try to alleviate the egocentricity of the epistemic cut.

\end{document}